\documentclass[prl,twocolumn,showpacs]{revtex4}
\usepackage{epsfig}
\usepackage{amsmath}

\begin{document}
\title{Induced quantum dots and wires: electron storage and delivery}
\author{S. Bednarek}
\author{B. Szafran}
\author{R.J. Dudek}
\author{K. Lis} \affiliation{Faculty of Physics and Applied
Computer Science, AGH University of Science and Technology, al.
Mickiewicza 30, 30-059 Krak\'ow, Poland}

\date{\today}

\begin{abstract}
We show that quantum dots and quantum wires are formed underneath
metal electrodes deposited on a planar semiconductor heterostructure
containing a quantum well. The confinement is due to the
self-focusing mechanism of an electron wave packet interacting with
the charge induced on the metal surface. Induced quantum wires guide
the transfer of electrons along metal paths and induced quantum dots
store the electrons in specific locations of the nanostructure.
Induced dots and wires can be useful for devices operating on the
electron spin.
\end{abstract}
\pacs{73.21.La,73.63.Nm} \maketitle

Planar nanodevices containing single \cite{0,1,2,3,4,5}, double
\cite{7,8,9,10}, and multiple \cite{11,11a} laterally coupled
quantum dots with confinement potential tuned by electrodes
deposited on top of the semiconductor heterostructure are at present
extensively studied in both theory and experiment in context of
application for quantum gates using electron spins as quantum bits.
Recent advances include demonstration that the electron spin can be
set and read-out \cite{1,3,4,5,6,7,8,9} as well as rotated
\cite{5,6}. In a quantum gate working on the electron spins
\cite{divi} the single qubit operations are to be performed with an
electron transfer to a high $g$ factor region or to a ferromagnetic
quantum dot where the electron spin is rotated by microwave
radiation. In this letter we present an idea for the control of the
electron localization and its transfer between specific locations
within the nanodevice. The idea is based on the self-focusing
mechanism of an electron wave packet near a conductor surface
\cite{18,19,20} which as we show below allows to exclude scattering
during the electron transfer and warrants the electron delivery to a
specific location in the device with a 100\% probability.

In the conventional planar nanodevices \cite{0,1} a negative
potential is applied to the gate electrodes to deplete the
two-dimensional electron gas underneath. In the variant of the
structure proposed below the role of the electrodes is different: a
single quantum-well-confined electron becomes self-trapped below the
conductor by the potential of the charge that it induces on the
metal surface. The response potential of the electron gas of the
conductor contains a component of lateral confinement which
localizes the quantum-well-confined electron in form of a wave
packet that moves parallel to the metal preserving its shape as an
electron soliton \cite{18,19}. The packed was called \cite{20} an
{\it inducton} since the focusing potential stems from the charge
induced in the electron gas. The inducton possesses mixed quantum
and classical properties. It is described by a wave function of both
spatial and spin coordinates whose time evolution is described by
the Schroedinger equation. On the other hand the inducton moves as a
stable wave packet of a finite size and its transition probability
in transport through potential barriers is binary (0 or 1)
\cite{19}.

\begin{figure}[ht!]
\includegraphics[bb = 38 360 577 834,
width=5cm]{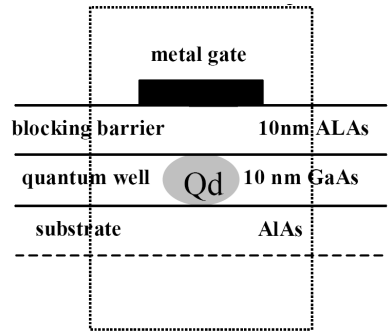}
 \caption{Cross section of a nanodevice generating an induced quantum dot or
 wire. Dotted line shows the boundaries of the computational box.}
 \label{f0}
\end{figure}

\begin{figure}[ht!]
  \includegraphics[bb = 56 190 562 634,
width=8cm]{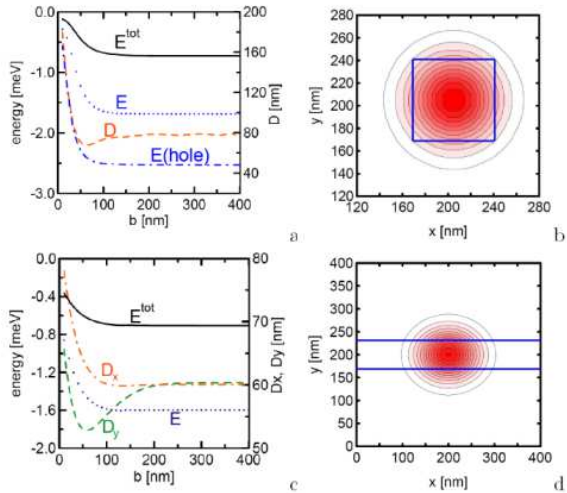}
 \caption{
 (a) Total energy $E^{tot}$, single electron energy $E$, and
 diameter of the wave packet $D$ for a
quantum dot induced by square metal plate of side length $b$.
Dash-dotted curve shows the eigenvalue $E$ for a heavy hole inducton
($m=0.45m_0$). (b) Electron density in the quantum dot induced by
square metal plate with $b=70$ nm. (c) $E^{tot}$, $E$ and packet
length along $x$ and $y$ direction ($D_x,D_y$) for a quantum wire
induced by an infinite metal bar of width $b$. (d) Charge density of
the wave packet confined under the metal bar of width $b=50$ nm.
 \label{f1}}
\end{figure}

\begin{figure}[ht!]
 \includegraphics[bb = 0 0 572 261,
width=5cm]{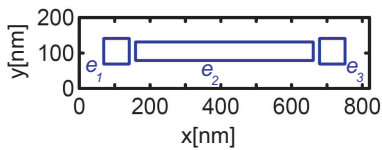}
 \caption{ Top view of the setup to transfer the
electron from the quantum dot induced under the electrode $e_1$ to
the quantum dot induced under $e_3$ via the quantum wire induced
under $e_2$.}
 \label{f2a}
\end{figure}

In Ref. \cite{19} we discussed a structure of planar infinite layers
of metal, insulator (or semiconducting blocking barrier) and a
quantum well in which the inducton was formed. Due to the
translational symmetry the inducton can be formed at any place under
an infinite metal plate and travel in any direction within the
quantum well. In this letter we show that a metal electrode of a
finite size is also able to trap an electron underneath it. For the
size of the electrode comparable to the inducton radius the
transverse motion of the packet is frozen and the induced charge
creates a confinement potential similar to the quantum dot
potential. A rectangular metal electrode of length larger than the
size of the self-focused wave packet leaves the inducton a single
degree of freedom for motion along the metal path which therefore
forms an induced quantum wire within the quantum well.

 The induced
potential calculated within the quantum linear response theory is
well approximated by the response of an ideal (classical) conductor
\cite{19,20}. Therefore, the induced potential can be quite
accurately evaluated by the classical electrodynamics. Let us
consider a nanodevice presented in Fig. \ref{f0}. For an infinite
metal plate the induced potential can be evaluated with the image
charge method \cite{19} which greatly simplifies the calculations
but is no longer applicable for metal plates of finite size. The
induced potential is therefore calculated from the Poisson equation.
We apply the theory used previously for the electrostatic quantum
dot modeling \cite{14} which describes experiments with remarkable
accuracy. The presence of the metal introduces fixed potential value
in the boundary conditions at the conducting surface. All the
nanodevice is contained in a rectangular computational box (see the
dotted line in Fig. \ref{f0}). We require the normal component of
the electric field at the surface of the box to vanish. The size of
the box is taken large enough that for an infinite metal plate we
reproduce the image charge potential. We solve the Poisson equation
\begin{equation}\nabla^2 \Phi({\bf r})=-\frac{1}{\epsilon
\epsilon_0} \rho({\bf r})\end{equation} where the charge density is
expressed by the electron wave function $\psi({\bf r})$ and the
electron charge $-e$:
\begin{equation}\rho({\bf r})=-e|\psi({\bf r})|^2.\end{equation}
According to the superposition principle  the calculated total
potential $\Phi$ is a sum of contributions stemming from two sources
\begin{equation}\Phi({\bf r})=\phi_1({\bf r})+\phi_2({\bf r}),\end{equation}
where $\phi_1$ is directly due to the charge density distribution
\begin{equation}
\phi_1({\bf r})=\frac{-e}{4\pi\epsilon\epsilon_0}\int d{\bf
r}'\frac{\rho({\bf r}')}{|{\bf r}-{\bf r}'|},
\end{equation}
and the second component is due to the charge induced on the
electrode which creates the lateral confinement for the electron
localized in the quantum well. Given the total potential $\Phi$ and
the potential of the electron packet $\phi_1$ the induced potential
$\phi_2$ is calculated according to
\begin{equation}\phi_2({\bf r})=\Phi({\bf r})-\phi_1({\bf
r}).\end{equation} For the nanostructure of Fig. \ref{f0} the motion
of the electron in the growth direction is frozen by the strong
vertical confinement which can be eliminated from the Schroedinger
equation thus taking a two dimensional form
\begin{equation}H=-\frac{\hbar^2}{2m}\left(\frac{\partial^2}{\partial x^2}+\frac{\partial^2}{\partial y^2}\right)-e\phi_2(x,y,z_0),\end{equation}
where $z_0$ is the center of the quantum well. In the eigenequation
\begin{equation}H\psi=E\psi,\end{equation}  $E$ is the
single-electron energy \cite{19}. The total energy is obtained
\cite{19} by subtracting half of the interaction energy of the
inducton with the induced potential:
\begin{equation}E^{tot}=E+\frac{e}{2}\int dxdy |\psi(x,y)|^2\phi_2(x,y,z_0).\end{equation}
The $\phi_2$ potential and the wave function $\psi$ are mutually
dependent so the problem is solved in a self-consistent iteration.

Let us assume that the metal electrode deposited on top of the
semiconductor (see Fig. \ref{f0}) is of a square shape with side
length $b$. We solve equations (1-8) and evaluate the average
diameter of the electron wave packet
\begin{equation}
D=2\int dxdy\sqrt{(x-x_0)^2+(y-y_0)^2}|\psi(x,y)|^2,
\end{equation}
where $(x_0,y_0)$ are the coordinates of the center of the metal
square. GaAs electron effective mass $m=0.067m_0$ and dielectric
constant $\epsilon=12.5$ are adopted.

The single-electron energy, the total energy and the packet diameter
are plotted as function of $b$ in Fig. \ref{f1}(a). Both the
energies are negative (for any $b$) and decrease with growing $b$
reaching the free inducton (electron wave packet under an infinite
metal plate) limit \cite{19}. The limit is obtained for $b$ larger
than the free inducton radius. The packet diameter is a non
monotonic function of $b$. For small plate the packet is large,
localization is the strongest for $b_{min}^{dot}=70$ nm and for
$b>b_{min}^{dot}$ the diameter grows to the free inducton limit.
Soon after the minimal diameter is reached the energies saturate as
function of $b$. Note that the minimal diameter is nearly equal to
$b_{min}^{dot}$ (for the charge density and the size of the plate
see Fig. \ref{f1}(b)). The $b_{min}^{dot}$ value is optimal for the
proposed applications of electron storage and transfer (see below).

\begin{figure}[ht!]
 \includegraphics[bb = 155 278 560
560, width=8cm]{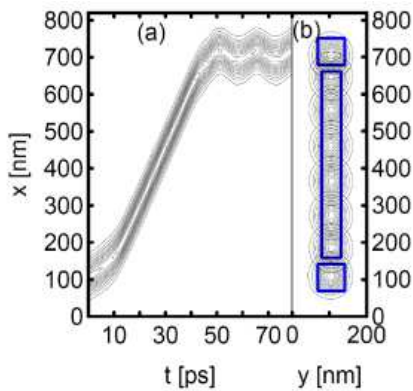}
 \caption{Time evolution of the electron density in the device with straight
 path.
 (a) Electron charge density as function of the $x$-variable and time calculated at the symmetry axis of the
 electrode configuration presented with blue lines in (b). The contour plots in (b) display the charge density
 at subsequent moments in time. }
 \label{f2}
\end{figure}

Similar calculation was performed for the electrode in form of a
metal bar that is infinite in the $x$-direction and of width $b$ in
the $y$-direction (it will be referred to as a current path). Fig.
\ref{f1}(c) shows both the energies and the size of the packet in
both directions of the quantum wire
\begin{equation}
D_x=2\int dxdy|x-x_0| |\psi(x,y)|^2
\end{equation}
\begin{equation}
D_y=2\int dxdy|y-y_0| |\psi(x,y)|^2.
\end{equation}
Quite remarkably values of $D_x$ and $D_y$  are close, although only
$D_y$ has a minimum as a function of $b$. The strongest focusing
appears for the current path of width $b_{min}^{wire}=50$ nm,
adopted as optimal in the following. The charge distribution for
$b=b_{min}^{wire}$ plotted in Fig. \ref{f1}(d) shows that the packet
is more strongly localized in the direction perpendicular to the
path.

\begin{figure}[ht!]

\begin{tabular}{c}
\includegraphics[bb = 138 83 450 767, width=4.75cm]{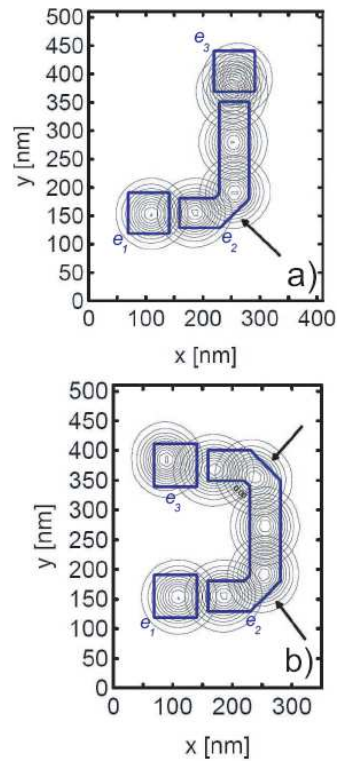}\end{tabular}

 \caption{Snapshots of the time evolution of the electron density following the path broken
once (a) or twice (b). The arrows indicate the cut corners of the
metal paths. The electron leaves the quantum dot induced under the
$e_1$ electrode and goes to the quantum dot under $e_3$.} \label{f3}
\end{figure}
In the calculations discussed above the stationary eigenproblem (7)
of Hamiltonian (6) was solved by the evolution in the imaginary time
\cite{22} which leads to formation of the ground state wave
function. A slight modification of the approach allows to
investigate the evolution in the real time. For that purpose the
time dependent Schroedinger equation is solved:
\begin{equation}
d\psi({\bf r},t)=\frac{i}{\hbar}H({\bf r},t)\psi({\bf r},t)dt.
\end{equation}
The dependence of the Hamiltonian on time appears for a moving
inducton due to the time dependence of the electron density which
enters Eqs. (1) and (4)
\begin{equation}
\rho({\bf r},t)=-e|\psi({\bf r},t)|^2.
\end{equation}
The time dependence of the potential is accounted  for by solving
Eqs. (1,4,5) for each time step (12).

Let us consider a nanodevice whose cross section agrees with the
schematic of Fig. \ref{f0} and the top view is displayed in Fig.
\ref{f2a}. On the surface of the structure we have three electrodes
separated by gaps of 20 nm. Quantum dots are formed below square
electrodes $e_1$ and $e_3$ both of size 70 nm $\times$ 70 nm. The
middle electrode $e_2$ (50 nm $\times$ 500 nm) is supposed to induce
a quantum wire which should guide an electron from under $e_1$ to
$e_3$.

We assume that the electron is confined below $e_1$ for a time long
enough to relax to the ground state. In the simulation it is
achieved by the imaginary time evolution for potentials of the
electrodes (the Schottky barier neglected): $V_1=0.0$ mV,
$V_2=V_3=-0.1$ mV ($V_1,V_2,V_3$ are the potentials applied to
$e_1,e_2,e_3$ electrodes respectively). After the inducton relaxes
to the ground state we change the applied voltages putting
$V_1=-0.1$ mV, $V_2= 0.0$ mV, $V_3=-0.1$ mV and we start the
evolution in the real time. The inversed $V_1-V_2$ potential
difference gently sets the electron in motion. The electron gains
the kinetic energy passing from below $e_1$ to $e_2$. Next it goes
along $e_2$ with a constant kinetic energy eventually reaching the
quantum dot induced under $e_3$. At that moment $V_2$ voltage is
switched to $V_2=-0.15$ mV to confine the electron permanently under
$e_3$. The traveling electron density is presented in Fig.
\ref{f2}(b) for several moments in time. In Fig. \ref{f2}(a)
$|\psi(x,y_0,t)|^2$ is plotted for $y_0=200$ nm set at the symmetry
axis of the proposed setup. We can see that the inducton is
accelerated between $e_1$ and $e_2$ (see the curved shape of the
density plot in Fig. \ref{f2}(a) between $x=100$ and 200 nm). Then
it moves with a constant velocity under $e_2$ [note increased
localization of the packet when under $e_2$]. Finally the packet
gets under $e_3$ and is trapped there when $V_2$ is switched to
negative. The oscillations observed in Fig. \ref{f2} (a) for $t>50$
ps are due to the residual kinetic energy which is not lost when the
inducton is trapped under $e_3$.  The crucial point of the presented
results is that the electron was transferred from one dot to the
other with a 100 \% probability which is due to the self-focusing
mechanism.

Let us consider a similar structures but with varied shape of the
electrodes (blue lines in Fig. \ref{f3}(a) and Fig. \ref{f3}(b)).
The middle electrode turns under right angles to force the inducton
to change direction of its velocity vector. The time evolution is
presented in Fig. \ref{f3}. It turns out that the electron can be
guided under any place in the structure also along curved paths.
Note the cut corner edges of the current path marked in Fig.
\ref{f3} with arrows. It allows the electron to change the motion
direction with equal incident and reflection angles. For an uncut
edge with a $90^\circ$ angle the electron is reflected back to
$e_1$.

The $b_{min}$ values adopted above for the size of the electrodes
are optimal for three reasons:  1) the confinement in quantum dots
is the strongest 2) the motion along the wires follows the axis of
the wire most closely 3) the electron still penetrates the region
that is not covered by the electrode that induces the confinement
[see Figs. \ref{f1}(c) and (d)] and can therefore be set in motion
by voltage applied to adjacent electrodes.

To conclude, we presented a design for a planar semiconductor
structure with a quantum well and electrodes separated by a tunnel
barrier in which induced quantum dots and quantum wires are formed.
The dots store the electrons in specific points of the nanostructure
and the paths assist in the transport of the electrons between
chosen locations in the device. The self-focusing effect allows the
electron to be kept in the stable wave packet (inducton state) of a
size close to the electrode width. It also allows the electron
transport with a 100\% probability of passing through obstacles in
form of potential cavities or barriers. The combination of
semi-classical transport properties of the inducton with its spin
degree of freedom is likely to become useful for the spin operating
devices. The inducton binding energy is not large (of order of 1.5
meV), but this is enough for the wave packet to overcome the
potential barriers on its way under the metal paths \cite{hole}.

\acknowledgments This work was supported by the State Committee for
Scientific Research  (KBN) under Grant No. 1P03B 002 27.


\begin{thebibliography}{00}
\bibitem{hole}
The binding energy of a hole inducton with $m=0.45m_0$ (see the
dash-dotted line in Fig. \ref{f1}(a)) is by about 1 meV stronger
since the larger mass allows for stronger localization which
enhances the electron-metal interaction energy.
\end{thebibliography}
\end{document}